

\font\titlefont = cmr10 scaled\magstep 4
\font\sectionfont = cmr10
\font\littlefont = cmr5
\font\eightrm = cmr8

\magnification = 1200

\global\baselineskip = 1.2\baselineskip
\global\parskip = 4pt plus 0.3pt
\global\abovedisplayskip = 18pt plus3pt minus9pt
\global\belowdisplayskip = 18pt plus3pt minus9pt
\global\abovedisplayshortskip = 6pt plus3pt
\global\belowdisplayshortskip = 6pt plus3pt


\def\endignore{}
\def\ignore #1\endignore{}

\newcount\dflag
\dflag = 0


\def\monthname{\ifcase\month
\or Jan \or Feb \or Mar \or Apr \or May \or June%
\or July \or Aug \or Sept \or Oct \or Nov \or Dec
\fi}

\def\timestring{{\count0 = \time%
\divide\count0 by 60%
\count2 = \count0
\count4 = \time%
\multiply\count0 by 60%
\advance\count4 by -\count0
\ifnum\count4 < 10 \toks1 = {0}
\else \toks1 = {} \fi%
\ifnum\count2 < 12 \toks0 = {a.m.}
\else \toks0 = {p.m.}
\advance\count2 by -12%
\fi%
\ifnum\count2 = 0 \count2 = 12 \fi
\number\count2 : \the\toks1 \number\count4%
\thinspace \the\toks0}}



\def\endtitle{}
\def\title#1\endtitle{\vskip.5in\titlefont
\global\baselineskip = 2\baselineskip
#1\vskip.4in
\baselineskip = 0.5\baselineskip\rm}

\def\endauthors{}
\def\authors#1\endauthors{#1}

\def\endabstract{}
\def\abstract#1\endabstract{\vskip .3in%
\centerline{\sectionfont\bf Abstract}%
\vskip .1in
\noindent#1}

\newcount\nsection
\newcount\nsubsection

\def\section#1{\global\advance\nsection by 1
\nsubsection=0
\bigskip\noindent\centerline{\sectionfont \bf \number\nsection.\ #1}
\bigskip\rm\nobreak}

\def\subsection#1{\global\advance\nsubsection by 1
\bigskip\noindent\sectionfont \sl \number\nsection.\number\nsubsection)\
#1\bigskip\rm\nobreak}

\def\topic#1{{\medskip\noindent $\bullet$ \it #1:}}
\def\endtopic{\medskip}

\def\appendix#1#2{\bigskip\noindent%
\centerline{\sectionfont \bf Appendix #1.\ #2}
\bigskip\rm\nobreak}


\newcount\nref
\global\nref = 1

\def\ref#1#2{\xdef #1{[\number\nref]}
\ifnum\nref = 1\global\xdef\therefs{\noindent[\number\nref] #2\ }
\else
\global\xdef\oldrefs{\therefs}
\global\xdef\therefs{\oldrefs\vskip.1in\noindent[\number\nref] #2\ }%
\fi%
\global\advance\nref by 1
}

\def\listrefs{\vfill\eject\section{References}\therefs}


\newcount\cflag
\newcount\nequation
\global\nequation = 1
\def\eqlabel{(1)}

\def\nexteqno{\ifnum\cflag = 0
\global\advance\nequation by 1
\fi
\global\cflag = 0
\xdef\eqlabel{(\number\nequation)}}

\def\lasteqno{\global\advance\nequation by -1
\xdef\eqlabel{(\number\nequation)}}

\def\label#1{\xdef #1{(\number\nequation)}
\ifnum\dflag = 1
{\escapechar = -1
\xdef\draftname{\littlefont\string#1}}
\fi}

\def\clabel#1#2{\xdef\eqlabel{(\number\nequation #2)}
\global\cflag = 1
\xdef #1{\eqlabel}
\ifnum\dflag = 1
{\escapechar = -1
\xdef\draftname{\string#1}}
\fi}

\def\cclabel#1#2{\xdef\eqlabel{#2)}
\global\cflag = 1
\xdef #1{\eqlabel}
\ifnum\dflag = 1
{\escapechar = -1
\xdef\draftname{\string#1}}
\fi}


\def\eeq{}

\def\eqnn #1\eeq{$$ #1 $$}

\def\eq #1\eeq{\xdef\draftname{\ }
$$ #1
\eqno{\eqlabel \rlap{\ \draftname}} $$
\nexteqno}



\def\eol{& \eqlabel \rlap{\ \draftname} \crcr
\nexteqno
\xdef\draftname{\ }}

\def\eeol{& \eqlabel \rlap{\ \draftname}
\nexteqno
\xdef\draftname{\ }}

\def\eolnn{\cr
\global\cflag = 0
\xdef\draftname{\ }}


\def\eqa #1\eeq{\xdef\draftname{\ }
$$ \eqalignno{ #1 } $$
\global\cflag = 0}


\def\ie{{\it i.e.\/}}

\def\etal{{\it et.al.\/}}
\def\apriori{{\it a priori\/}}
\def\aposteriori{{\it a posteriori\/}}
\def\via{{\it via\/}}


\def\anp#1#2#3{{\it Ann. Phys. (NY)} {\bf #1} (19#2) #3}
\def\arnps#1#2#3{{\it Ann.~Rev.~Nucl.~Part.~Sci.} {\bf #1}, (19#2) #3}
\def\ijmp#1#2#3{{\it Int. J. Mod. Phys.} {\bf A#1} (19#2) #3}

\def\mpla#1#2#3{{\it Mod.~Phys.~Lett.} {\bf A#1}, (19#2) #3}

\def\npb#1#2#3{{\it Nucl. Phys.} {\bf B#1} (19#2) #3}
\def\plb#1#2#3{{\it Phys. Lett.} {\bf #1B} (19#2) #3}

\def\prd#1#2#3{{\it Phys. Rev.} {\bf D#1} (19#2) #3}
\def\pr#1#2#3{{\it Phys. Rev.} {\bf #1} (19#2) #3}

\def\prl#1#2#3{{\it Phys. Rev. Lett.} {\bf #1} (19#2) #3}

\def\zpc#1#2#3{{\it Zeit. Phys.} {\bf C#1} (19#2) #3}


\global\nulldelimiterspace = 0pt



\def\frac#1#2{{{#1} \over {#2}}\,}  
\def\hf{{1\over 2}}



\def\Dsl{\hbox{/\kern-.6000em\it D}} 
\def\dsl{\hbox{/\kern-.5600em$\partial$}}
\def\pxpsl{\hbox{/\kern-.5600em$p$}}
\def\ssl{\hbox{/\kern-.5600em$s$}}
\def\epssl{\hbox{/\kern-.5600em$\epsilon$}}
\def\delsl{\hbox{/\kern-.7000em$\nabla$}}
\def\lxpsl{\hbox{/\kern-.5600em$l$}}
\def\kxpsl{\hbox{/\kern-.5600em$k$}}
\def\qxpsl{\hbox{/\kern-.5600em$q$}}
\def\sla#1{\raise.15ex\hbox{$/$}\kern-.57em #1}



\def\roughly#1{\mathrel{\raise.3ex\hbox{$#1$\kern-.75em\lower1ex\hbox{$\sim$}}}}

\def\ol#1{\overline{#1}}




\def\Bfw{{\bf W}}


\def\Sca{{\cal A}}

\def\Scd{{\cal D}}

\def\Scl{{\cal L}}

\def\Sct{{\cal T}}

\def\Scw{{\cal W}}

\def\Scz{{\cal Z}}


\def\tr{\mathop{\rm tr}}
\def\Tr{\mathop{\rm Tr}}

\def\Det{\mathop{\rm Det}}
\def\Log{\mathop{\rm Log}}







\ref\bernreuther{W. Bernreuther, U. L\"ow, J.P. Ma and O. Nachtmann,
\zpc{43}{89}{117}.}

\ref\fermionsnonlin{R.D. Peccei, S. Peris and X. Zhang, \npb{349}{91}{305}.}

\ref\fermionsnoninv{G. Valencia and A. Soni, \plb{263}{91}{517}.}

\ref\fermionslin{
W. Buchm\"uller and D. Wyler, \npb{268}{86}{621};
J. Anglin, C.P. Burgess, H. de Guise, C. Mangin and J.A. Robinson,
\prd{43}{91}{703};
C.P. Burgess and J.A. Robinson, \ijmp{6}{91}{2707};
A. De R\'ujula, M.B. Gavela, O. P\`ene and F.J. Vegas, \npb{357}{91}{311};
M. Traseira and F.J. Vegas, \plb{262}{91}{12}.}

\ref\wwv{
K.J.F. Gaemers and G.J. Gounaris, \zpc{1}{79}{259};
K. Hagiwara, R.D. Peccei, D. Zeppenfeld and K. Hikasa, \npb{282}{87}{253}.}

\ref\gbtreenonlin{
C.L. Bilchak and J.D. Stroughair, \prd{30}{84}{1881};
J.A. Robinson and T.G. Rizzo, \prd{33}{86}{2608};
S.-C. Lee and W.-C. Su, \prd{38}{88}{2305}; \plb{205}{88}{569};
\plb{212}{88}{113}; \plb{214}{88}{276};
C.-H. Chang and S.-C. Lee, \prd{37}{88}{101};
D. Zeppenfeld and S. Willenbrock, \prd{37}{88}{1775};
U. Baur and D. Zeppenfeld, \npb{308}{88}{127}; \plb{201}{88}{383};
\npb{325}{89}{253}; \prd{41}{90}{1476};
G. Couture, S. Godfrey and P. Kalyniak, \prd{39}{89}{3239};
K. Hagiwara, J. Woodside and D. Zeppenfeld, \prd{41}{90}{2113};
V. Barger and T. Han, \plb{241}{90}{127};
E.N. Argyres, O. Korakianitis, C.G. Papadopoulos and W.J. Stirling,
\plb{259}{91}{195};
E.N. Argyres and C.G. Papadopoulos, \plb{263}{91}{298};
S. Godfrey, Carleton preprint OCIP-C-91-2;
S. Godfrey and H. Konig, Carleton preprint OCIP-C-91-5;
G. Couture, S. Godfrey and R. Lewis, \prd{45}{92}{777}.}

\ref\gbloopnonlin{
S.J. Brodsky and J.D. Sullivan, \pr{156}{67}{1644};
F. Herzog, \plb{148}{84}{355}; \plb{155}{85}{468}E;
M. Suzuki, \plb{153}{85}{289};
J.C. Wallet, \prd{32}{85}{813};
A. Grau and J.A. Grifols, \plb{154}{85}{283}; \plb{166}{86}{233};
J.J. van der Bij, \prd{35}{87}{1088};
F. Hoogeveen, Max Planck Inst. preprint MPI.PAE/PTh 25/87 (1987),
unpublished.
W.J. Marciano and A. Queijeiro, \prd{33}{86}{3449};
G.L. Kane, J. Vidal and C.-P. Yuan, \prd{39}{89}{2617};
G. B\'elanger, F. Boudjema and D. London, \prl{665}{90}{2943};
F. Boudjema, K. Hagiwara, C. Hamzaoui and K. Numata, \prd{43}{91}{2223}.}

\ref\gblooplin{
D. Atwood, C.P. Burgess, C. Hamzaoui, B. Irwin and J.A. Robinson,
\prd{42}{90}{3770};
F. Boudjema, C.P. Burgess, C. Hamzaoui and J.A. Robinson,
\prd{43}{91}{3683}.}

\ref\criticisms{A. de R\'ujula, M.B. Gavela, P. Hernandez and E. Mass\'o,
CERN preprint CERN-Th.6272/91, 1991.}

\ref\effectivelagrangian{S. Weinberg, Physica {\bf 96A} (1979) 327;
J. Polchinski, \npb{231}{84}{269};
H. Georgi, {\it Weak Interactions and Modern Particle Theory}
(Benjamin/Cummings Menlo Park, 1984);
C.P. Burgess and J.A. Robinson, in {\it BNL Summer Study on CP Violation}
S. Dawson and A. Soni editors, (World Scientific, Singapore, 1991).
}

\ref\lorentzinvariant{
R. K\"ogerler and D. Schildknecht, CERN preprint CERN-TH.3231 (1982),
unpublished;
D. Schildknecht, in "Electroweak effects at high energies", ed.
N.B. Newman (Plenum, New York, 1985) p. 551.;
M. Kuroda, D. Schildknecht and K.-H. Schwarzer, \npb{261}{85}{432};
J. Maalampi, D. Schildknecht, and K.-H. Schwarzer, \plb{166}{86}{361};
H. Neufeld, J.D. Stroughair and D. Schildknecht, \plb{198}{87}{563};
M. Kuroda, J. Maalampi, D. Schildknecht and K.-H. Schwarzer,
\plb{190}{87}{217}; \npb{284}{87}{271};
M. Kuroda, F.M. Renard and D. Schildknecht, \plb{183}{87}{366};
C. Bilchak, M. Kuroda and D. Schildknecht, \npb{299}{88}{7};
Y. Nir, \plb{209}{88}{523};
J.A. Grifols, S. Peris and J. Sol\`a, \plb{197}{87}{437}; \ijmp{3}{88}{225};
R. Alcorta, J.A. Grifols and S. Peris, \mpla{2}{87}{23};
C. Bilchak and J.D. Stroughair, \prd{41}{90}{2233};
H. Schlereth, \plb{256}{91}{267}.}

\ref\nonlinearrealization{J.M. Cornwall, D.N. Levin and G. Tiktopoulos,
\prd{10}{74}{1145};
B.W. Lee, C. Quigg and H. Thacker, \prd{16}{77}{1519};
M. Veltman, Acta. Phys. Pol. {\bf B8} (1977) 475;
M.S. Chanowitz and M.K. Gaillard, \npb{261}{85}{379};
M.S. Chanowitz, M. Golden and H. Georgi, \prd{36}{87}{1490};
M.S. Chanowitz, \arnps{38}{88}{323};
R.D. Peccei and X. Zhang, \npb{337}{90}{269};
B. Holdom and J. Terning, \plb{247}{90}{88};
J. Bagger, S. Dawson and G. Valencia, preprint BNL-45782;
M. Golden and L. Randall, \npb{361}{91}{3};
B. Holdom, \plb{259}{91}{329};
A. Dobado, M.J. Herrero and D. Espriu, \plb{255}{91}{405};
R.D. Peccei and S. Peris, \prd{44}{91}{809};
A. Dobado and M.J. Herrero, preprint CERN-TH-6272/91.
}

\ref\ccwz{C. Callan, S. Coleman, J. Wess and B. Zumino, \pr{177}{69}{2239};
\pr{177}{69}{2247};
J. Gasser and H. Leutwyler, \anp{158}{84}{142}.}

\ref\cutoff{C.P. Burgess and David London, preprint McGill-92/05,
UdeM-LPN-TH-84.}


\def\gwk{$SU_L(2) \times U_Y(1)$}
\def\gem{$U_{\rm em}(1)$}
\def\em{{\rm em}}
\def\ngb{Nambu-Goldstone boson}
\def\zdm{$Z$dm}


\rightline{March 1992}
\rightline{McGill-92/04}
\rightline{UdeM-LPN-TH-83}
\vskip .3in

\title
\centerline{Light Spin-One Particles}
\centerline{Imply Gauge Invariance}
\endtitle

\authors
\centerline{C.P. Burgess${}^a$ and David London${}^b$\footnote{}{email:
cliff@physics.mcgill.ca; london@lps.umontreal.ca}}
\vskip .15in
\centerline{\it ${}^a$ Physics Department, McGill University}
\centerline{\it 3600 University St., Montr\'eal, Qu\'ebec, CANADA, H3A 2T8.}
\vskip .1in
\centerline{\it ${}^b$ Laboratoire de Physique Nucl\'eaire, Universit\'e de
Montr\'eal}
\centerline{\it C.P. 6128, Montr\'eal, Qu\'ebec, CANADA, H3C 3J7.}
\endauthors

\abstract
Recently, calculations which consider the implications of anomalous trilinear
gauge-boson couplings, both at tree-level and in loop-induced processes, have
been criticized on the grounds that the lagrangians employed are not \gwk gauge
invariant. We prove that, in fact, the general Lorentz-invariant and $U(1)_\em$
invariant but {\it not} $SU_L(2)\times U_Y(1)$ invariant action is
equivalent to the general lagrangian in which $SU_L(2)\times U_Y(1)$
appears but is nonlinearly realized. We demonstrate this equivalence in
an explicit calculation, and show how it is reconciled with loop
calculations in which the different formulations can (superficially) appear
to give different answers. In this sense any effective theory containing
light spin-one particles is seen to be automatically gauge invariant.
\endabstract


\section{Introduction}

Perhaps the biggest pothole in the otherwise reasonably well-maintained
surface that is high-energy theory is our ignorance of the origin of
particle masses. This ignorance is patched over in the standard model
through the introduction of the Higgs couplings, but a better understanding
is expected once shorter distance scales have been probed. One way in which
the physics underlying the Higgs sector might make itself known in
accelerator collisions is through the deviations from standard-model
predictions it can produce in the couplings of particles to gauge-boson
probes. The couplings of the massive $W^\pm$ and $Z^0$ bosons themselves
are  particularly interesting in this regard since they directly involve
the symmetry-breaking physics through their longitudinal modes.

This type of reasoning has led to considerable effort in outlining the
potential form that the anomalous couplings of these particles to the
photon and the $Z^0$ might take, since these are the probes that are
currently the most cleanly available in collider experiments. Since the
experimental success of the standard model up to and beyond the $Z^0$ mass
can be interpreted as saying that the energy scale appropriate to any new
physics must be large, the analysis of potential anomalous couplings has
focused on the lowest electromagnetic and electroweak moments of fermions
\bernreuther, \fermionsnonlin, \fermionsnoninv, \fermionslin\ and gauge
bosons \wwv, \gbtreenonlin, \gbloopnonlin, \gblooplin, \criticisms\ that
would dominate interactions at low energies. The natural theoretical
framework for this type of analysis is an effective-lagrangian approach
\effectivelagrangian\ in which the influence of any at-present-unknown new
heavy particles is parameterized through the effective nonrenormalizable
interactions that they generate among the lighter particles.

Essentially only two ingredients are required to specify such a low-energy
effective lagrangian: the low-energy particle content and the symmetries
that their interactions preserve. Although by and large there is agreement
on the low-energy  particle content, there are currently two main choices
that  are made concerning the symmetries that should be required of the
low-energy lagrangian. One school \bernreuther, \fermionsnoninv,
\gbtreenonlin, \gbloopnonlin, \lorentzinvariant\ imposes only the very
minimal conditions of Lorentz-invariance, $SO(3,1)$, and electromagnetic
gauge invariance, \gem. The alternative procedure \fermionsnonlin,
\nonlinearrealization\ is to require invariance with respect to the full
electroweak gauge group, $SU_L(2) \times U_Y(1)$, but with all but the
unbroken \gem\ subgroup being nonlinearly
realized.\footnote{${}^\dagger$}{\eightrm A third choice \fermionslin,
\gblooplin, \criticisms\ is to linearly realize \gwk-invariance by
explicitly including the standard-model Higgs doublet in the low-energy
theory. We do not pursue this option further here.}  In this second
framework the unknown symmetry-breaking sector is assumed at low energies
to contain only the three Nambu-Goldstone bosons which are eaten by the
massive $W^\pm$ and $Z^0$ particles. The transformation properties of all
fields are then determined by general arguments \ccwz\ that were developed
within the framework of chiral perturbation theory many years ago.

The principal goal of this article is to demonstrate the equivalence of
these two schemes. We show that each may be obtained from the other \via\
field redefinitions. This demonstration is given in section (2) below. A
practical implication of this equivalence is to permit the application of
renormalizable gauges to loop calculations in what is nominally not a
gauge-invariant theory. At a conceptual level it illustrates that
spontaneously broken gauge invariance is automatic for {\it any} effective
theory containing light spin-one particles.

Recently, de R\'ujula and coworkers have criticized most analyses involving
anomalous trilinear gauge-boson couplings, saying that the lagrangians used
are not gauge invariant \criticisms. Although many of their conclusions are
basically correct, the equivalence theorem established in this paper shows
that the supposed non-gauge invariance of such effective lagrangians is
actually a red herring. Incorrect conclusions that are based on the
non-gauge invariant effective lagrangian really arise from other abuses of
the effective-lagrangian formalism. We pursue these related issues in a
separate publication \cutoff.

Of course, the claimed equivalence only makes sense within the domain of
applicability of both formulations of the effective theory. For both
approaches this is necessarily restricted to energy scales that are not too
large compared to the spin-one boson masses, $M$. For a weak spin-one
coupling, $g$, the maximum applicable scale may be estimated to be $\simeq
4 \pi M/g$ in order of magnitude. At higher energies pathologies such as
the failure of perturbative unitarity may be expected, indicating a
breakdown of the low-energy approximation and the appearance of some sort
of `new physics'.

In order to bring out some of the peripheral issues which can confuse this
equivalence we compute the one-loop-induced weak fermion dipole moment that
would be generated by a particular (anapole) anomalous moment in the $WWZ$
interaction. We show that although the equivalence is manifest within a
gauge-invariant regularization---such as dimensional regularization---it is
hidden when a cutoff is used (as is frequently done in the literature). In
this case the induced weak dipole moment can be quadratically or
logarithmically divergent depending on the gauge, or even on the field
variables that are  employed.

Although some of these points are undoubtedly known to the
effective-lagrangian cognicenti, it is evident
that they have not percolated out into the wider community which is now
finding applications for these techniques. (This is particularly clear in
criticisms \criticisms\ of the `non-gauge invariant' formulation discussed
above.)
For this reason we feel that a re-examination of these issues is
appropriate here.

\section{The Equivalence Result}

There are two natural ways to incorporate spontaneously broken gauge
symmetries within a low-energy effective lagrangian:

\topic{No Gauge Invariance}
In the first formulation massive spin-one bosons are represented by vector
fields and the lagrangian is only required to be Lorentz invariant. Only
invariance with respect to unbroken gauge symmetries is imposed and
all broken gauge symmetries are simply ignored.

\topic{Nonlinearly-Realized Gauge Invariance}
The alternative is the second approach in which both Lorentz and gauge
symmetries are built in from the beginning. Spontaneous symmetry breaking
is incorporated by coupling all fields to a symmetry-breaking sector. All
that is assumed about this sector is that its only light degrees of freedom
are the appropriate set of Nambu-Goldstone bosons that are required on
general grounds by Goldstone's theorem. These are, of course, ultimately
`eaten' by the gauge bosons \via\ the Higgs mechanism once the
nonlinearly-realized action of the broken-symmetry transformations amongst
the Nambu-Goldstone bosons is `gauged'.
\endtopic

We demonstrate in this section a precise form for the equivalence of these
two formulations for the low-energy lagrangian.
Although the arguments can be made quite generally, we restrict ourselves
here to establishing this equivalence for two specific cases: a simplified
toy model involving a single massive spin-one particle, as well as the
realistic case appropriate to the couplings of the electroweak gauge
bosons, $W^\pm$, $Z^0$ and the photon, $\gamma$.

\subsection{A Toy Example}

In order to describe the argument within its simplest context, consider
first the coupling of a single massive spin-one particle, $V_\mu$, coupled
to various forms of spinless or spin-half matter, $\psi$. We first state
the two alternative forms for the effective lagrangian and then demonstrate
their equivalence.

\topic{No Gauge Invariance}
The lagrangian in the first formulation then takes the form:
\eq
\label\lagrangone
\Scl_1 = \Scl_1(V_\mu,\psi),
\eeq
in which $\Scl_1$ is \apriori\ an arbitrary local Lorentz-invariant
function of the fields $V_\mu$, $\psi$ and their spacetime derivatives.
Since $\psi$ and $V_\mu$ are independent degrees of freedom the quantum
theory could be defined in this case by a functional integral of the form:
\eq
\label\fintone
Z_1 = \int [d\psi] \; [dV_\mu] \; \exp\left[ i\int d^4x \; \Scl_1
(V_\mu,\psi) \right].
\eeq

\topic{Nonlinearly Realized Gauge Invariance}
The alternative formulation is to consider a $U(1)$ gauge theory with
matter fields, $\chi_i$, carrying $U(1)$ charges $q_i$. The gauge symmetry
transformations acting on these fields and on the gauge potential, $A_\mu$,
are the usual ones:
\eq
\chi_i \to e^{iq_i \omega} \; \chi_i; \qquad gA_\mu \to gA_\mu +
\partial_\mu \omega.
\eeq
$g$ here is the gauge coupling constant.

Symmetry breaking is incorporated by coupling these matter and gauge fields
in a completely general way to a single \ngb, $\varphi$, for a
spontaneously broken $U(1)$. The action of the $U(1)$ on the \ngb s may
always be chosen to take a  standard form \ccwz, which becomes in this case
\eq
\varphi \to \varphi + f \omega.
\eeq
$f$ here is the \ngb's decay constant which is of the order of the scale at
which the $U(1)$ symmetry is spontaneously broken. It is related to the
mass  of the gauge boson by the relation $M = gf$.

The most general gauge-invariant low-energy lagrangian may then be written
in the following form:
\eq
\label\lagrangtwo
\Scl_2 = \Scl_2(D_\mu\varphi,\chi'),
\eeq
in which the redefined field is $\chi'_i \equiv e^{-i q_i \varphi/f} \;
\chi_i$ and the gauge-covariant derivative for $\varphi$ is given by $D_\mu
\varphi \equiv \partial_\mu \varphi - g f A_\mu$. Notice that all of the
dependence on $A_\mu$ in $\Scl_2$ arises through this gauge-covariant
derivative. For example, the gauge field strength is given by $gf F_{\mu\nu}
= \partial_\mu D_\nu \varphi - \partial_\nu D_\mu \varphi$.

The corresponding functional integral defining the quantum theory then has
the standard form:
\eq
\label\finttwo
Z_2 = \int [d\chi'_i] \; [dA_\mu] \; [d\varphi] \; \exp\left[ i\int
d^4x \; \Scl_2 (D_\mu\varphi,\chi') \right] \; \delta[G] \;
\Det\left(\frac{\delta
G}{\delta \omega} \right),
\eeq
in which the second-to-last term is the functional delta function,
$\delta[G]$, which enforces the gauge condition $G = 0$, and the last term
is the associated Fadeev-Popov-DeWitt---or ghost---functional determinant.

It is crucial for the remainder of the argument that both $\chi'_i$ and
$D_\mu \varphi$ are {\it in}variant---as opposed to being {\it
co}variant---with respect to gauge transformations. As a result even if the
lagrangian, $\Scl_2$, is only required to be Lorentz invariant it becomes
automatically also gauge invariant.

\topic{Equivalence}
Now comes the main point. The two lagrangians, $\Scl_1$ and $\Scl_2$, are
identical to one another. There is a one-to-one correspondence between the
terms in each given by the replacement $\psi \leftrightarrow \chi'_i$ and
$D_\mu \varphi \leftrightarrow - gf \, V_\mu$. This is only possible
because {\it both} $\Scl_1$ and $\Scl_2$ are constrained only by Lorentz
invariance and so any interaction which is allowed for one is equally
allowed for the other.

More formally, the functional integral of eq. \fintone\ may be obtained
from that of eq. \finttwo\ by simply choosing unitary gauge, defined by the
condition $G \equiv \varphi(x)$, and using the functional delta function to
perform the integration over $\varphi$. The ghost `operator' is in this
case $\delta G(x)/\delta \omega(x') = f \; \delta^4(x-x')$ and so the ghost
determinant contributes just a trivial field-independent normalization
factor.

The integration over the `extra' Nambu-Goldstone degree of freedom of the
gauge-invariant theory is thereby seen to be precisely compensated by the
freedom to choose a gauge.

\subsection{Applications to the Electroweak Bosons}

The argument as applied to a more complicated symmetry-breaking pattern,
such as appears in the electroweak interactions, has essentially the same
logic although the technical details are slightly more intricate.

\topic{No Gauge Invariance}
We take for the purposes of illustration the degrees of freedom in the
low-energy effective lagrangian for the electroweak interactions of
leptons and quarks. These are: the massless photon, $A_\mu$, the massive weak
vector bosons, $W_\mu$ and $Z_\mu$, and the usual fermions, $\psi$. Although
other particles such as gluons may also be very simply included we do not do so
here for simplicity of notation. The general lagrangian for these fields may be
written:
\eq
\label\smlagrangone
\Scl_1 = \Scl_1(A_\mu,W_\mu, Z_\mu,\psi),
\eeq
in which $\Scl_1$ is a general local and Lorentz-invariant function whose
form is constrained only by the requirement of invariance with respect to
the unbroken electromagnetic gauge transformations, \gem. All
derivatives are taken to be the \gem\ gauge-covariant derivative,
$D_\mu$, which for fermions takes the form $D_\mu \psi = \partial_\mu \psi -ie
Q
A_\mu \psi$. $Q$ here denotes the diagonal matrix of fermion electric charges.

The quantum theory is given in terms of a functional integral of the form
\eq
\label\lowbrowint
Z_1 = \int [dW_\mu] \; [dW^*_\mu] \; [dZ_\mu] \; [dA_\mu] \;
[d\ell_i] \; \exp\left[ i\int d^4x \; \Scl_1 \right] \;
\delta\left[G_\em \right] \; \Det\left(\frac{\delta G_\em}{\delta
\omega_\em} \right).
\eeq

We next outline the nonlinear realization of \gwk.

\topic{Nonlinearly Realized Gauge Invariance}
The first step is to briefly review the formulation for realizing
the symmetry-breaking pattern $SU_L(2) \times U_Y(1) \to U_\em(1)$ nonlinearly
\ccwz.

Consider, therefore, a collection of matter fields, $\psi$, on which \gwk\ is
represented (usually reducibly) by the matrices $G = \exp[
i\omega_2^a T_a + i \omega_1 Y]$. We choose here a slightly unconventional
normalization for the generators $T_a$ and $Y$, {\it viz} $\tr[ T_a T_b] = \hf
\, \delta_{ab}$, $\tr[T_a Y] = 0$ and $\tr[Y^2] = \hf$. Finally define the
matrix-valued scalar field containing the Nambu-Goldstone bosons by $\xi(x) =
\exp[i X_a \varphi^a(x) /f]$, in which the three $X_a$'s represent the
spontaneously broken generators $X_1 = T_1$, $X_2 = T_2$ and $X_3 = T_3 - Y$.
$X_3$ here is chosen to be orthogonal to the unbroken generator of \gem:
$Q = T_3 + Y$.

The action of the gauge group \gwk\ on $\xi$ and $\psi$ may be written in the
standard form \ccwz:
\eq
\label\realization
\psi \to G \psi \quad \hbox{and} \quad
\xi \to \xi', \quad \hbox{where} \quad G\; \xi = \xi' \; H^\dagger.
\eeq
Here $H = \exp[i Q \, u(\xi,\xi',G)]$ and $u = u(\xi,\xi',G)$ is implicitly
defined by the condition that $\xi'$ on the right-hand-side of eq.
\realization\
involves only the broken generators.

As was the case for the toy example, for the purposes of constructing the
lagrangian it is convenient to define new matter fields, $\psi'$,
according to $\psi' \equiv \xi^\dagger \; \psi$ since this has the \gwk\
transformation rule:
\eqa
\psi' &\to \xi'^\dagger \; G \, \psi \eolnn
&= H \; \psi'.  \eeol
\eeq
Notice that even for global $U_Y(1)$ rotations, for which $\omega_1$ is
constant, $u(\xi,\xi',G)$ is spacetime dependent because of its dependence on
the scalar field $\xi(x)$.

The next step is the construction of the general locally \gwk\ invariant
effective lagrangian. To this end consider the auxiliary quantity
$\Scd_\mu(\xi)$ which may be defined in terms of $\xi$ and the \gwk\ gauge
potentials $\Bfw_\mu = g_2 W^a_\mu \, T_a + g_1 B_\mu \, Y$ by
\eq
\label\auxiliary
\Scd_\mu(\xi) \equiv \xi^\dagger \partial_\mu\xi -i \xi^\dagger \Bfw_\mu \xi.
\eeq
In terms of this quantity it is possible to construct fields which transform
in a simple way with respect to \gwk. Together with their transformation rules
these are,
\eqa
\label\connections
e\, \Sca_\mu & \equiv i \, \tr[ Q \Scd_\mu(\xi)], \qquad e\Sca_\mu \to
e\Sca_\mu
+ \partial_\mu u; \eol
\sqrt{g_1^2 + g_2^2} \; \Scz_\mu &\equiv 2i \, \tr[(T_3 - Y)
\Scd_\mu(\xi)],\qquad \Scz_\mu \to \Scz_\mu; \eol
g_2 \, \Scw^\pm_\mu &\equiv i\sqrt{2} \, \tr[T_\mp \Scd_\mu(\xi)], \qquad
\Scw^\pm_\mu \to e^{\pm iu Q} \, \Scw_\mu^\pm. \eeol
\eeq
$T_\pm$ is defined as usual to be $T_1 \pm iT_2$. The first of these
fields, $\Sca_\mu(\xi)$, transforms in such a way as to permit the construction
of a covariant derivative for the local transformations as realized on $\psi'$:
\eq
\label\covderiv
D_\mu \psi' \equiv (\partial_\mu - i e\Sca_\mu \, Q )\; \psi'.
\eeq

The main point to be appreciated here is that all of the fields $\psi'$,
$D_\mu\psi'$, $\Sca_\mu(\xi)$, $\Scz_\mu(\xi)$ and $\Scw_\mu^\pm(\xi)$
transform
purely electromagnetically under arbitrary \gwk\ transformations. This ensures
that once the lagrangian is constructed to be invariant under the unbroken
group,
\gem, it is {\it automatically} invariant with respect to the full
nonlinearly-realized group \gwk.

With these transformation rules the most general \gwk-invariant lagrangian
becomes
\eq
\Scl_2 = \Scl_2(\Sca_\mu,\Scw_\mu,\Scz_\mu,\psi')
\eeq
with $\Scl_2$ restricted only by the unbroken \gem\ gauge
invariance. The functional integral which defines the quantum theory may
then be written
\eq
\label\holygrailint
Z_2 = \int [d\Bfw_\mu] \; [d\xi] \; [d\psi'] \; \exp\left[
i\int d^4x \; \Scl_2 \right] \; \delta\left[G_a \right] \;
\Det\left(\frac{\delta G_a}{\delta \omega^b} \right).
\eeq
Four gauge conditions, $G_a = 0$, $a=1,...4$, are required---one for each
generator of \gwk.

\topic{Equivalence}
The demonstration of the equivalence between eqs. \lowbrowint\ and
\holygrailint\ proceeds along lines that are similar to those used in the
abelian toy example presented previously. As was the case in this earlier
example, the equivalence works term-by-term in the lagrangian. The
correspondence between the field variables is
\eq
\label\correspond
\Sca_\mu \leftrightarrow A_\mu, \quad
\Scz_\mu \leftrightarrow  Z_\mu, \quad
\Scw_\mu^\pm \leftrightarrow W^\pm_\mu, \quad
\psi' \leftrightarrow \psi.
\eeq

The equivalence is explicit in unitary gauge, which is defined in this case
by the condition $\varphi^a(x) \equiv 0$, or equivalently $\xi(x) \equiv 1$,
throughout spacetime. As is seen from the transformation rules of eq.
\realization\ this condition does not completely fix the gauge. It is preserved
by the unbroken electromagnetic transformations which satisfy $G = H =
e^{i\omega_{\rm em}}$. In this gauge the relations for $\Scz_\mu$, $\Scw_\mu$
and $\psi$ indicated in eqs. \correspond\ above simply become equalities.

More formally, using the unitary gauge-condition to perform the functional
integral over $\xi$ in eq. \holygrailint, gives the result
\eq
\label\holygrailtwo
Z_2 = \int [d\Bfw_\mu] \; [d\psi] \;  \exp\left[
i\int d^4x \; \Scl_2 \right] \; \delta\left[G_\em \right] \;
\Det\left(\frac{\delta G_\em}{\delta \omega_\em} \right)  \;
\left.\Det \left( \frac{\delta\varphi^a}{\delta\omega^b} \right)
\right|_{\varphi=0}.
\eeq
Since $\Scl_2(\xi =1) = \Scl_1$ this clearly agrees with eq. \lowbrowint\
apart from the final Fadeev-Popov-DeWitt ghost determinant that is
associated with the choice of unitary gauge
\eq
\delta\varphi^a(x)/\delta \omega^b(x') \equiv {\Delta^a}_b(x) \;
\delta^4(x-x').
\eeq

The final point is that the identity $\Det \equiv \exp \Tr \Log$ may be used
to rewrite this determinant as the exponential of a local, Lorentz- and
$U_{\rm em}(1)$-invariant function. As such it may be considered as a shift
in the parameters appearing in the original lagrangian, $\Scl_2$.
Furthermore, since  its contribution to $\Scl_2$ is proportional to
$\delta^4(x=0)$ its coefficients  are ultraviolet divergent and so their
contribution may be absorbed into the renormalizations that are anyhow
required in defining the functional integral of eq. \holygrailtwo. At a
practical level, the Fadeev-Popov determinant does not in any case arise until
at least two-loop order.

\section{An Illustrative Calculation}

In order to illustrate explicitly the equivalence of the two formulations,
we will compute the $CP$-violating `weak dipole moment' \bernreuther\ (which
we denote by $Z$dm) of the $\tau$ lepton,
\eq
\label\zdmdef
\Scl_{\rm zdm} = - iz \; \ol{\tau} \; \gamma_5 \sigma^{\mu\nu} \tau \;
\partial_\mu Z_\nu,
\eeq
that is induced at one loop by an anomalous $WWZ$ vertex. We consider for
these purposes the following $CP$-violating anomalous anapole coupling such
as appears in the non-gauge invariant formulation of Hagiwara \etal\
ref.~\wwv:\footnote{${}^\dagger$}{ \eightrm The coefficient `$a$' in
this equation corresponds to $g_4^Z$ of ref.~\wwv.}
\eq
\label\anapole
\Scl_a = - a \; W_\mu^* W_\nu  \left(\partial^\mu Z^\nu + \partial^\nu
Z^\mu\right).
\eeq

We may translate this effective interaction into a form in which the gauge
invariance is nonlinearly realized using the general correspondence of the
previous section. The result is to simply make the substitutions of
eqs.~\correspond\ in eq.~\anapole.

In order to illustrate the equivalence of these two formulations we next
compute the \zdm\ using the anapole vertex as derived from interaction
\anapole\ before and after making the substitution \correspond.

\subsection{Unitary Gauge Calculation}

In the non-gauge-invariant formulation the anapole vertex of Fig.~1 is
represented by the following Feynman rule
\eq
\label\anapolerule
a \left( k^\beta g^{\mu\alpha} + k^\alpha g^{\mu\beta} \right),
\eeq
and the gauge bosons propagate with the usual massive vector-boson
propagator
\eqa
\label\mvprop
G^{\mu\nu}_U(k) &= -i \frac{P^{\mu\nu}(k)}{k^2 - M_W^2} \eolnn
{\rm with}\quad P^{\mu\nu}(k) &= g^{\mu\nu} - \frac{k^\mu k^\nu}{M_W^2}.
\eeol
\eeq

The expression for $z$ may then be read from the amplitude (see Fig.~2)
\eqa
\label\ampunitary
\Sct^\mu = - {a g_w^2\over 2}  \int {d^n q \over \left(2\pi\right)^n} \;
{1\over D} & \left( k^\beta g^{\mu\alpha} + k^\alpha g^{\mu\beta} \right)
P_{\alpha\rho}( q + p_2) \; P_{\beta\sigma}(q - p_1) \eolnn
& ~~~~~~{\overline{u}}_\tau\left(p_2\right) \gamma^\rho \qxpsl
\gamma^\sigma \gamma_L \; v_\tau\left(p_1\right), \eeol
\eeq
$D$ here represents the denominators of the propagators that appear in the
graph
\eq
D = (q^2 - m^2_{\nu_\tau}) \left[(q+p_2)^2 - M_W^2\right]
\left[ (q-p_1)^2 - M_W^2\right].
\eeq
Since this amplitude diverges we regularize the integral by working in $n
\ne 4$ dimensions. We will return to the issue of regularization later in
this section. The divergent part may be explicitly evaluated to be
\eq
\label\answerdimreg
\Sct^\mu = - {ag_w^2\over 384\pi^2} \thinspace
{m_\tau \left(m_\tau^2 - m^2_{\nu_\tau}\right) \over M_W^4} \thinspace
{\overline{u}}_\tau\left(p_2\right) \sigma^{\mu\nu} k_\nu \gamma_5
v_\tau\left(p_1\right) \left({2\over 4-n}\right),
\eeq
and may be absorbed by renormalizing the coefficient $z$ of the
\zdm\ operator of eq.~\zdmdef. This determines how these operators mix due
to renormalization. In the minimal subtraction scheme we therefore find:
\eq
\label\running
z(\mu) = z(\mu') + {g_w^2\over 384\pi^2} \thinspace
{m_\tau \left(m_\tau^2 - m^2_{\nu_\tau}\right) \over M_W^4} \; a(\mu') \;
\log\left(  \frac{\mu^2}{\mu'^2} \right).
\eeq

\subsection{Renormalizable-Gauge Calculation}

The same calculation may be performed in a general gauge using the Feynman
rules appropriate to the effective lagrangian with nonlinearly-realized
gauge invariance. The principal difference here is that there are now four
diagrams -- that of Fig.~2, and those in which one or both of the $W^\pm$'s is
replaced by the corresponding would-be-Goldstone boson (WBGB), $\varphi^\pm$.

In the standard family of covariant renormalizable gauges parameterized by
the variable $\alpha$ the $\varphi^\pm$-scalar and $W^\pm$-boson propagators
are
respectively given by
\eqa
G_{(\alpha)}(k) &= \frac{i}{k^2 - \alpha M_W^2} \eol
\label\proprelation
{\rm and}\quad G^{\mu\nu}_{(\alpha)}(k) &= -i \frac{1}{ k^2 - M^2} \;
\left[ g^{\mu\nu} + (\alpha -1) {k^\mu k^\nu\over k^2 - \alpha M_W^2 }
\right] \eolnn
&= G^{\mu\nu}_U(k)  - \frac{k^\mu k^\nu}{M_W^2} \; G_{(\alpha)}(k).  \eeol
\eeq

As is clear from the expansion of $\Scw_\mu(\xi)$ in terms of powers of
fields:
\eq
\Scw^\pm_\mu = g_2 \left[ W^\pm_\mu  + \frac{1}{M_W}
\partial_\mu \varphi^\pm + \cdots \right],
\eeq
the Feynman rule for the emission of a WBGB, $w$, of four-momentum $k^\mu$
from the anapole vertex is found by simply contracting the result for the
emission of the corresponding gauge particle---\ie\ that of
eq.~\anapolerule---by $k^\mu/M_W$. The same is true for the emission of a
WBGB by a fermion line. As may be easily verified these are precisely the
vertices that are required to preserve the $\alpha$-independence of tree
level amplitudes.

{}From these Feynman rules it is immediately clear that the sum of the four
graphs that contribute in the renormalizable gauges precisely corresponds
to the four terms that would be obtained by substituting eq.~\proprelation\
into the unitary-gauge result of eq.~\ampunitary. This demonstrates the
equivalence of the induced \zdm\ as computed with the non-gauge-invariant
and the nonlinearly-realized gauge-invariant formulations.

Notice that this equivalence has relied on the WBGB's having derivative
couplings to fermions as well as to the anapole vertex. Such couplings are
an  automatic consequence of the replacement \correspond\ in the
nonlinearly-realized effective lagrangian. They differ superficially from those
that appear in the standard model, however, where the WBGB's couple to fermions
\via\ renormalizable Yukawa couplings. This difference is irrelevant because
one
set of couplings may be changed into the other by performing an appropriate
field redefinition, which cannot alter any scattering amplitudes. It is in fact
straightforward to check that use of these Yukawa couplings in the previous
calculation does not at all alter our conclusions.

\subsection{Related Red Herrings}

This equivalence as outlined appears to be so simple as to be almost
trivial. It is therefore worth outlining some circumstances which can act,
and have acted in the literature, to obscure this conclusion.

The main obstacle to understanding this equivalence is the widespread use
of cutoffs to regularize the divergent integrals that arise in loop-level
effective-lagrangian applications. For the present purposes an uncritical
use of cutoffs can cause confusion in two distinct ways. At a purely
technical level they can hide the transformation properties of the theory
under field redefinitions in general, and gauge transformations in
particular, and so can give the impression of obtaining differing results
in different gauges. Cutoffs also introduce a more conceptual difficulty
once an attempt is made to associate a physical interpretation with the
cutoff-dependence of a given amplitude. We speak briefly to each of these
issues in the following paragraphs.

At the technical level, it is notoriously easy to inadvertently break
gauge-invariance with a cutoff regularization. One way to see this is to
implement the cutoff in the effective theory by adding higher-derivative
kinetic terms to the lagrangian. This has the effect of multiplying each
propagator by a form factor which separately implements the cutoff on each
internal line of any graph and ensures, for example, that the cutoff result
is independent of extraneous issues such as how momentum is routed through
the graph. Considered this way, however, it is clear that higher-derivative
terms cannot be gauge invariant unless the derivatives used are gauge
covariant. Gauge covariant derivatives necessarily imply additional
cutoff-dependent interaction terms, however, whose effects are easily
missed if cutoffs are simply applied \aposteriori\ to loop integrals.

A related issue concerns the behaviour of cutoff-regulated amplitudes under
field redefinitions. For instance, in the example considered above it is
superficially possible to change the divergent behaviour of the result
simply by performing a field redefinition. This may be seen by comparing
the result of evaluating the given graph using two kinds of WBGB--fermion
couplings: on the one hand using the derivative WBGB--fermion couplings
which come from the general substitution \correspond, and on the other hand
using the standard-model Yukawa-type couplings between these particles.

In order to see these difficulties explicitly consider using the following
form factor regularization in the one-loop-generated \zdm
\eq
\label\regulator
{-\Lambda^2 \over q^2 - \Lambda^2}
{}~~{-\Lambda^2 \over \left(q+p_2\right)^2 - \Lambda^2 }
{}~~{-\Lambda^2 \over \left(q-p_1\right)^2 - \Lambda^2 }.
\eeq
Using this regularization together with the derivatively-coupled
fermion--WBGB vertex one finds the following quadratic divergence
\eq
\label\answerunitary
\Sct^\mu = - {a g_w^2\over 2304 \pi^2 } \;\frac{\Lambda^2}{M_W^4} m_\tau
\thinspace {\overline{u}}_\tau\left(p_2\right)
\sigma^{\mu\nu} k_\nu \gamma_5 v_\tau\left(p_1\right).
\eeq
This result holds for both the unitary-gauge and the $\alpha$-gauge
calculations.

Performing the same calculation using Yukawa-type WBGB--fermion vertices in
$\alpha$-gauge gives instead only linear and logarithmic divergences. These
arise only from the graph in which both vector bosons in Fig.~2 are
replaced by WBGB's. The result from this graph is
\eqa
\label\ampfeynman
\Sct^\mu = - {ag_w^2\over 2 M_W^4}
\int {d^4 q \over \left(2\pi\right)^4} \; {1\over D} &
\left[ 2 q^\mu q\cdot k - q \cdot k \left( p_1 - p_2 \right)^\mu \right]
\eolnn
& \qquad {\overline{u}}_\tau\left(p_2\right) \left[ \qxpsl \left( m_\tau^2
\gamma_R + m^2_{\nu_\tau} \gamma_L \right) - m_\tau m^2_{\nu_\tau} \right]
v_\tau\left(p_1\right).\eeol
\eeq
Regularizing using eq.~\regulator\ as before, we find
\eq
\label\answerfeynman
\Sct^\mu = - {a g_w^2 \over 384 \pi^2} \thinspace
{m_\tau \left(m_\tau^2 - m^2_{\nu_\tau}\right) \over M_W^4}
\; \ln{\Lambda^2 \over M_W^2 } \thinspace
{\overline{u}}_\tau\left(p_2\right) \sigma^{\mu\nu} k_\nu \gamma_5
v_\tau\left(p_1\right),
\eeq
which is only logarithmically divergent, as advertised.

The problem here is that these two kinds of Feynman rules for the
fermion--WBGB vertex may be obtained from one another by performing a
WBGB-dependent nonlinear field redefinition on the fermion fields. The
answer would be unchanged if the higher-derivative term which implements
the cutoff were also transformed, since this transformation would introduce
new cutoff-dependent fermion--WBGB interactions. Of course, this is not
what was compared between eqs. \answerunitary\ and \answerfeynman.

There are two lessons to be learned from this example. The first is that it
is very simple to miss contributions when performing field redefinitions on
cutoff-regulated quantities. More important, however, is the realization
that the cutoff dependence of an amplitude in an effective theory is not
necessarily simply related to its dependence on the heavy mass scales that
appear within whatever short-distance physics generates that effective
lagrangian. Since cutoffs are frequently used to estimate the scale of new
physics which might be probed in proposed experiments, we will deal with
this issue in more depth in a separate publication \cutoff. It suffices
here to remark that the connection between cutoffs and the scale of new
physics is completely unrelated to how gauge-invariance is realized in the
effective lagrangian. Furthermore, we repeat that superficial
gauge-variance of cutoff-regulated results can usually be traced to the
non-invariance of the regularization -- and not to the lagrangian itself.

\section{Conclusions}

Effective lagrangians are the natural way to parameterize the effects of
the new physics that is ultimately responsible for the breaking of the
electroweak gauge group. If one does not wish to explicitly include a Higgs
scalar in the low-energy theory, there are two principal candidates for
such an effective lagrangian -- one which requires only $U_\em(1)$ gauge
invariance, but not $SU_L(2)\times U_Y(1)$ gauge invariance, and one which
imposes the full $SU_L(2)\times U_Y(1)$ gauge invariance, nonlinearly
realized. We have demonstrated the equivalence of these two lagrangians.

The same arguments as are used here may be similarly used to prove this
equivalence for more general symmetry-breaking patterns $G \to H$. This
shows that any effective theory containing light spin-one particles
automatically has a (spontaneously broken) gauge invariance. Alternatively,
one can say that at low energies there is little to choose between a
spontaneously-broken gauge invariance and no gauge invariance at all. It
also shows that criticisms of effective lagrangians based on the absence of
gauge invariance are actually red herrings. Problems with these lagrangians
tend to arise for other reasons, such as the careless use of cutoffs to
regularize loop diagrams.

At a practical level this equivalence has the advantage that it allows the
use of the techniques of renormalizable gauges for calculations in what is
nominally not a gauge-invariant theory. This is useful when powercounting
arguments are being used in that all propagators explicitly vary like
$1/p^2$ for large four-momenta. As a simple example, this equivalence
provides an extremely easy way to see why QED remains renormalizable even
after it is supplemented by a photon mass term while a nonabelian gauge
theory like the standard model does not. The difference may be most easily
seen in the version of these theories in which the WBGB's are explicit. It
arises because although it is possible to construct an invariant
power-counting renormalizable lagrangian for a $U(1)$ WBGB -- simply its
kinetic term $-\hf \, D_\mu \varphi D^\mu \varphi$ -- such a term is {\it
not} possible for a nonabelian symmetry group. This is because the kinetic
terms are in this case not by themselves invariant with respect to the
nonlinearly-realized symmetries.

\bigskip
\centerline{\bf Acknowledgments}
\bigskip

D.L. thanks F. del Aguila for the hospitality of the University of Granada,
where part of this work was done. Many thanks also to Fawzi Boudjema,
Steven Godfrey, Yossi Nir, Santi Peris, and Xerxes Tata and German Valencia for
helpful criticism, and to Markus Luty for pointing out an error in an earlier
draft. This research was partially funded by funds from the N.S.E.R.C.\ of
Canada and les Fonds F.C.A.R.\ du Qu\'ebec.

\vfill\eject
\centerline{\bf Figure Captions}
\bigskip

\topic{Figure 1}
The Feynman rule for the CP-violating anomalous gauge-boson vertex discussed
in the text. All momenta are outgoing.

\topic{Figure 2}
The Feynman graph through which the anomalous gauge-boson vertex
contributes to fermion weak dipole moments.

\listrefs

\bye